\pdfoutput=1
\documentclass[%
 reprint,
superscriptaddress,
 amsmath,amssymb,
 prl
]{revtex4-1}

\usepackage{graphicx}
\usepackage{dcolumn}
\usepackage{bm}
\usepackage{hyperref}

\usepackage{newtxtext}
\usepackage{xcolor}
\usepackage{braket}
\usepackage{tikz}

\newcommand{\cT}{\mathcal{T}}
\newcommand{\cB}{\mathcal{B}}
\newcommand{\cH}{\mathcal{H}}
\let\oldparagraph\paragraph
\renewcommand{\paragraph}[1]{\oldparagraph{#1---}\hspace{-1em}}

\begin{document}

\title{Anomaly Constraints on Gapped Phases with Discrete Chiral Symmetry}

\author{Clay C\'{o}rdova}
 \email{clayc@uchicago.edu}
\affiliation{Kadanoff Center for Theoretical Physics \& Enrico Fermi Institute, University of Chicago
}%

\author{Kantaro Ohmori}
 \email{komori@scgp.stonybrook.edu}
\affiliation{%
Simons Center for Geometry and Physics, SUNY
}%
\date{\today}

\begin{abstract}
    We prove that in $(3+1)d$ quantum field theories with $\mathbb{Z}_N$ symmetry, certain anomalies forbid a symmetry-preserving vacuum state with a gapped spectrum.   In particular, this applies to discrete chiral symmetries which are frequently present in gauge theories as we illustrate in examples.  Our results also constrain the long-distance behavior of certain condensed matter systems such as Weyl-semimetals and may have applications to crystallographic phases with symmetry protected topological order.  These results may be viewed as analogs of the Lieb-Schultz-Mattis theorem for continuum field theories.

\end{abstract}

\maketitle


\paragraph{Introduction} Symmetry is a universally applicable tool to constrain the dynamics of strongly-coupled quantum field theories.  At the most foundational level, a symmetry organizes energy eigenstates into representations of the symmetry group and provides selection rules constraining the dynamics.  

Beyond these elementary considerations there are more subtle aspects of symmetry in quantum field theory such as 't Hooft anomalies. These are mild violations of gauge invariance of the partition function of the theory coupled to classical background gauge field sources.  Such violations are characterized by local functionals and are intrinsic to the theory when they cannot be eliminated by a choice of local counterterms.  

One way to understand anomalies and their properties is via anomaly inflow \cite{Callan:1984sa}.  In this framework the anomaly of a theory is characterized by a local classical action for background gauge fields in one higher spacetime dimension.  These classical actions are sometimes called invertible field theories following \cite{Freed:2004yc}. The notion of locality invoked here means that the action obeys certain cutting and gluing rules and applies even for discrete background gauge fields appropriate for discrete global symmetries.  The classification of invertible field theories, and in turn anomalies, has a close connection with topology and cobordism theory \cite{KTalk1, Kapustin:2014tfa, Kapustin:2014gma, Kapustin:2014dxa, KTalk, Freed:2016rqq, Gaiotto:2017zba, Yonekura:2018ufj}.  In condensed matter physics, the long-distance limit of symmetry-protected topological order (SPT) is described by such a classical action. The anomalous field theory is then a set of edge modes which may reside on the boundary of the SPT.  

A crucial feature of anomalies is that they are invariant under any continuous deformation of a quantum field theory.
This includes changing the energy scale of observation so that anomalies are invariant under renormalization group flow \cite{tHooft:1979rat}.   Anomalies thus provide powerful input on the possible long-distance behavior of strongly-coupled quantum field theories.  

It is useful to organize the possible long-distance physics of a given quantum field theory by the behavior of the vacuum under the global symmetry and whether or not there exists a gap in the energy spectrum.  Anomalies computed in a weakly coupled ultraviolet description can then constraint which of these various behaviors is possible at long distances.  In particular, while it is believed that any anomaly can be carried by a symmetry breaking, or gapless vacuum, it is sometimes the case that a given anomaly is inconsistent with a gapped and symmetry-preserving vacuum.  The fluctuations around such a low energy phase are described by a topological quantum field theory so one can rephrase the above by stating that the possible anomalies of unitary topological field theories with a unique ground state on the sphere are restricted.  

Our main result below is to illustrate an example of such a restriction in the context of $(3+1)d$ quantum field theories with $\mathbb{Z}_{N}$ global symmetry.  This symmetry and anomaly occurs in a wide variety of theories, in particular in gauge theories with discrete chiral symmetry where our result implies that a mass gap necessitates discrete chiral symmetry breaking.

The fact that anomalies can prohibit a gap in the energy spectrum is in some cases well known.  For instance in $(3+1)d$ field theories, the most familiar cubic anomalies of continuous symmetry groups arising from triangle diagrams in perturbative theories have this feature.  One way to argue this is that such a diagram represents a portion of the three-point function of current operators with power law behavior at separated points and hence requires massless degrees of freedom \cite{Coleman:1982yg}.

By contrast, the implications of discrete anomalies, i.e.\ either discrete anomalies for continuous groups such as parity anomalies in $(2+1)d$, or anomalies for finite symmetry groups, is less clear.  In particular, these anomalies do not have any obvious imprint on the correlation functions of local operators and so the above logic does not apply. There are general constructions \cite{Witten:2016cio,Wang:2017loc,Wan:2018djl,Kobayashi:2019lep,Kobayashi:2019xxg} showing that large classes of anomalies for discrete symmetry groups can be matched by a suitable symmetry preserving topological field theory.  

Converse results have also been obtained.  Indeed, in certain cases \cite{Wang:2014lca, Wang:2016gqj, Sodemann:2016mib, Wang:2017txt} it has been argued that a given discrete anomaly is incompatible with a symmetry preserving gapped phase in $(2+1)d$ by investigating the physics of anyons, and analogous results have been found using higher-form symmetry \cite{Kobayashi:2018yuk}.  Recently \cite{Cordova:2019xxx} derived a general obstruction extracted from any anomaly, which must necessarily vanish if the anomaly admits a symmetry preserving gapped phase.  We now apply this to anomalies for $\mathbb{Z}_N$ symmetry in $(3+1)d$ theories.

\paragraph{The $\mathbb{Z}_N$ SPT from $(4+1)d$ Fermions} Let us describe the $\mathbb{Z}_{N}$ anomaly of interest.  An intuitive way to proceed is to first think of the $\mathbb{Z}_{N}$ as part of a larger $U(1)$ symmetry.  In that case, there are two distinct types of anomalies, a cubic $U(1)$ anomaly and a mixed anomaly between $U(1)$ and Poincar\'{e} symmetry (sometimes called a mixed $U(1)$-gravity anomaly).  When $U(1)$ is reduced to $\mathbb{Z}_{N}$ both types of anomalies survive (though the classification becomes somewhat complicated \cite{Garcia-Etxebarria:2018ajm,Hsieh:2018ifc}.)  Our focus is on the discrete analog of the mixed anomaly.  As we will see below, often $\mathbb{Z}_{N}$ symmetries with such an anomaly arise from chiral symmetries that act on fermions.  

More directly, we can summarize the $\mathbb{Z}_{N}$ anomaly by inflow from a $(4+1)d$ SPT protected by the same symmetry group.  
The SPT, which we call $\cT^k_N$, can be associated with the long-distance limit of a $k$ copies of a massive $(4+1)d$ fermion $\Psi$ with real mass term $m\overline{\Psi}\Psi$.  The $\mathbb{Z}_N$ symmetry acts on these fields as $\Psi\to e^{\frac{2\pi\mathrm{i}}{N}}\Psi$. In the limit $|m|\to \infty$ the theory becomes trivially gapped, but maintains subtle information about the symmetry which constrains the boundary physics.

We can see the imprint of the $\mathbb{Z}_N$ symmetry by taking space to be a compact smooth spin 4-manifold $M_4$ with minimal non-zero value, 48, of the Pontryagin number $\int_{M_4}p_1(TM_4)$.  On such a manifold, the Atiyah-Singer index theorem \cite{atiyah1963index} implies that the fermions have $2k$ complex zero modes $\lambda_i$, $i=1,\cdots 2k$.  Therefore, when the size of $M_{4}$ is small compared to the mass, the system is described by the $0+1d$ complex fermions $\lambda_i$ which are acted on by the $\mathbb{Z}_N$ symmetry.

Quantization proceeds in a standard fashion.  Let  $\ket{-}$ be the state annihilated by the operators $\lambda_i$: $\lambda_i\ket{-}=0$.  By convention we can take $\ket{-}$ to be invariant under the $\mathbb{Z}_N$ symmetry.  Now set $\ket{+}:=\prod_{i=1}^{2k}\lambda^\dag_i\ket{-}$. On this state the generator, $U$, of the $\mathbb{Z}_N$ symmetry acts as
\begin{equation}\label{plusact}
U\ket{+}=e^{-\frac{4k\pi\mathrm{i}}{N}}\ket{+}~.
\end{equation}
The effect of the $(4+1)d$ fermion mass $m$ is to break the degeneracy between these states and leave a unique vacuum.  We set conventions so that the energy of $\ket{+}$ is positively proportional to $m$.  Then, when $m$ is positive the system flows to the trivial phase, with a unique $\mathbb{Z}_{N}$ invariant ground state.  On the other hand, when $m$ is negative and $N \nmid 2k$, the ground state $\ket{+}$ is not invariant under the symmetry and thus defines a nontrivial SPT $\cT^k_N$.

From the construction above it is clear that this SPT always admits a simple boundary defined by $k$ massless $(3+1)d$ Weyl fermion. Such a boundary can be constructed by allowing the mass of the fermions $\Psi$ to modulate along the fifth dimension $x^5$: we set $m>0$ where $x^5<0$, $m< 0$ where $x^5> 0$, and $m=0$ at $x^5=0$. In the region $x^5< 0$ the system flows to the trivial phase, and in the region $x^5> 0$ the system flows to $\cT^k_N$. On the interface, $x^5=0,$ there are localized massless $(3+1)d$ Weyl fermions.

The possible fermionic $\mathbb{Z}_N$ SPTs in $(4+1)d$ were classified in \cite{Garcia-Etxebarria:2018ajm,Hsieh:2018ifc}.  Their result takes the from of $\mathbb{Z}_{a_N}\oplus \mathbb{Z}_{b_N}$ where $a_N$, $b_N$ are integers depending on $N$ whose explicit form is given in the references.  The SPT $\cT^1_N$ corresponds to a generator of this classification.  Our analysis above shows that if $N\mid 2k$ the phase $\cT^k_N$ is trivial on $M_4 \times S^1$
when the $\mathbb{Z}_{N}$ symmetry is not activated along $M_{4}$.  This does not in general mean that $\cT^k_N$ is trivial as an SPT.  However, it is known 
\cite{Garcia-Etxebarria:2018ajm,Hsieh:2018ifc} that if $N\mid 2k$, the phase $\cT^k_N$ admits a gapped symmetry-preserving boundary.  As we now argue, if $N \nmid 2k$ no such boundary exists \footnote{\cite{Hsieh:2018ifc} proved that the symmetry enhancement method does not work if $N\nmid 2k$. We do not assume any particular construction.}.

\paragraph{No Symmetry-Preserving Boundary Topological Order} We constrain the possible $(3+1)d$ boundary theories $\cB$ following the general logic of \cite{Cordova:2019xxx}.  Consider the SPT $\cT^k_N$ on $M_4\times \mathbb{R}_+$, where $M_4$ is a compact smooth spin-4-manifold with Pontryagin number 48 as above, and $\mathbb{R}_+ = \{x^5|x^5\ge 0\}$ is the half-line.  The theory $\cB$ lives on the boundary $x^5=0$.

The path-integral on the region $0\le x^5 \le x$ with $x>0$ defines a boundary state $\ket{\cB}$ in the Hilbert space of $\cT^k_N$ on $M_4\times \{x\}$.  From our discussion above, we know that this Hilbert space has dimension one and is spanned by $\ket{+}$.  Therefore, $\ket{\cB} = Z_{\cB}[M_4]\ket{+}$ where the coefficient $Z_{\cB}[M_4]$ should be regarded as the partition function of the theory $\cB$ on $M_4$. If the boundary $\cB$ does not explicitly break the $\mathbb{Z}_N$ symmetry, the boundary state satisfies $U\ket{\cB}=\ket{\cB}$. However, from \eqref{plusact} we know that  $U\ket{+} \neq \ket{+}$.  Therefore we conlude
\begin{equation}\label{eq:ZcBnull}
    Z_{\cB}[M_4] = 0~.
\end{equation}
When $\cB$ is the $(3+1)d$ massless Weyl fermion, this equation is satisfied because of the fermion zero modes.

Now we assume that the boundary is a unitary spin-topological field theory $\cB_\text{top}$.  Since we are only interested in symmetry-preserving boundaries, we further assume that the Hilbert space of states, $\cH_{S^3}$, of $\cB_\text{top}$ on a sphere $S^{3}$ is one-dimensional. We argue that these assumptions are incompatible with \eqref{eq:ZcBnull}. 

Define the state $\ket{0}\in \cH_{S^3}$ by the path-integral over the disk $D^4$. The $S^4$ partition function is the norm: $Z_{\cB_\text{top}}[S^4] = \braket{0|0}$ which is positive by unitarity.  More generally, given a connected compact 4-manifold $X_{4}$, we define a state $\ket{X_{4}}\in \cH_{S^3}$ by the path-integral over $X_{4}\setminus D^4$, where $D^4$ is a small ball around a point in $X_{4}$.  The inner product $\braket{X_{4}|Y_{4}}$ of two of such states is identified with the partition function $Z_{\cB_\text{top}}[X_{4}\#Y_{4}]$ on the connected sum $X_{4}\#Y_{4}$.  Noting also that $\frac{\ket{0}\bra{0}}{\braket{0|0}}$ is the unit operator on $\cH_{S^3}$, we have
\begin{equation}\label{eq:Zconnected}
    Z_{\cB_\text{top}}[X_{4}\#Y_{4}] = Z_{\cB_\text{top}}[X_{4}]Z_{\cB_\text{top}}[Y_{4}]Z_{\cB_\text{top}}[S^4]^{-1}~.
\end{equation}

Next we make use of \cite{wall1964simply}, which implies that:
given a simply-connected smooth spin 4-manifold $X_{4}$, there exists an integer $\ell$ such that 
\begin{equation}\label{wall}
    X_{4}\#(-X_{4})\# (S^2\times S^2)^{\#\ell}  \stackrel{\text{diff.}}{\cong}(S^2\times S^2)^{\#(\ell+\chi(X_{4})-2)}~,
\end{equation}
where $-X_{4}$ is the orientation reverse of $X_{4}$, $(S^2\times S^2)^{\#\ell}$ is the connected sum of $\ell$ copies of $S^2\times S^2$, and $\chi(X_{4})$ is the Euler number of $X_{4}.$ In general, the partition function on $X_{4}$ and $-X_{4}$ are complex conjugates.  Hence \eqref{eq:Zconnected} and \eqref{wall} imply that for a simply connected spin manifold $X_{4},$ the absolute value of the partition function depends only on its Euler number:
\begin{equation}\label{eq:Zsimpconn}
    |Z_{\cB_\text{top}}[X_{4}]|^2 = Z_{\cB_\text{top}}[S^2\times S^2]^{\chi(X_{4})-2}Z_{\cB_\text{top}}[S^4]^{4-\chi(X_{4})}.
\end{equation}

Finally, following  \cite{Cordova:2019xxx} we prove that 
\begin{equation}\label{eq:ZSS}
    Z_{\cB_\text{top}}[S^2\times S^2]\neq 0~.
\end{equation} 
The path-integral over $D^2\times S^2$ defines a state $\ket{\phi}$ in the Hilbert space on $S^2\times S^1$, and $Z_{\cB_\text{top}}[S^2\times S^2]=\braket{\phi|\phi}$. As $S^4$ can be obtained by gluing $D^2\times S^2$ and $S^1 \times D^3$ along their boundary $S^1\times S^2$, we have $Z_{\cB_\text{top}}[S^4]=\braket{\phi|\tilde{\phi}}$ where the state $\tilde{\phi}$ is defined by path-integral over $S^1\times D^3$.  Since $Z_{\cB_\text{top}}[S^4]>0$ from unitarity
we conclude $\ket{\phi}\neq 0$ and thus $Z_{\cB_\text{top}}[S^2\times S^2]=\braket{\phi|\phi}\neq 0$.

Therefore, \eqref{eq:Zsimpconn}-\eqref{eq:ZSS} imply that the partition function on a simply-connected $X_{4},$ of any unitary spin topological field theory with a unique state on $S^{3}$ obeys
\begin{equation}
    Z_{\cB_\text{top}}[X_{4}] \neq 0~.
\end{equation}
In particular, taking $X_{4}$ to have Pontryagin number 48, e.g.\ the K3 manifold, this contradicts \eqref{eq:ZcBnull}.

To summarize, we have shown that the $(4+1)d$ SPT $\cT^k_N$ associated with $k$ $(4+1)d$ massive fermions charged under the $\mathbb{Z}_N$ symmetry cannot have a symmetry-preserving gapped boundary when $N\nmid 2k$. If the boundary is gapped, the symmetry must be broken down to $\mathbb{Z}_{\mathrm{gcd}(N,2k)}$ at least.  On the way, we have also proven that the partition function of a $(3+1)d$ spin-topological field theory on a simply-connected manifold obeys the simple relation \eqref{eq:Zsimpconn}.

\paragraph{Applications to Gauge Theories} Here we apply our results to the chiral $\mathbb{Z}_N$ symmetries in a gauge theory with simple gauge group $G$ and $N_f$ copies of massless Weyl fermions $\psi$ in a representation $R$. If necessary for gauge anomaly cancellation, we also add $N_f$ Weyl fermions in the complex conjugate representation $\overline{R}$.

We consider the chiral symmetry acting on $\psi$ (and not on $\overline{\psi}$) simultaneously.  This classical $U(1)$ symmetry is explicitly broken because of the ABJ anomaly.  However, depending on $R$, a discrete chiral symmetry group   remains:
\begin{equation}
U(1)\longrightarrow \mathbb{Z}_{N_f I(R)}~,
\end{equation}
where  $I(R)$ is the index of the representation $R$ \footnote{We define $I(R)$ as $\mathrm{Tr}_{R}T^aT^b = \frac{I(R)}{2h^\vee_{G}}\delta^{ab}$ with $h^\vee_G$ the dual Coxeter number and $T^a$ normalized as $\mathrm{Tr}_{\mathbf{adj}}T^aT^b = \delta^{ab}$.}.

This discrete chiral $\mathbb{Z}_{N_f I(R)}$ symmetry often has the type of anomaly discussed in the previous sections. (The appropriate $k$ above is simply the total number of fermions.) We thus deduce that such theories cannot be gapped without discrete chiral symmetry breaking when $I(R)\nmid 2\mathrm{dim}(R)$.  When this obstruction occurs in examples is summarized in Table~\ref{tab:obst}.

\begin{table}[t]
    \begin{tabular}{cccccc}
        $G$ & $R$ & $I(R)$ & $\mathrm{dim}(R)$ & Obstruction? & Enforced vacua\\
        \hline
        $SU(N_c)$ & $\mathbf{fund}$ & 1 & $N_c$ & No & 1\\
        $SO(N_c)$ & $\mathbf{vec}$ & 2 & $N_c$ & No& 1\\
        $SU(N_c)$ & $\mathbf{adj}$ & $2N_c$ & $N_c^2-1$ & Yes & $N_c$\\
        $SU(N_c)$ & $\mathbf{sym}$ & $N_c+2$ & $\frac{N_c(N_c+1)}{2}$ & Yes &  $\frac{N_c}{2}$ or $N_c$\\
        $SU(N_c)$ & $\mathbf{asym}$ & $N_c-2$ & $\frac{N_c(N_c-1)}{2}$ & Yes &  $\frac{N_c}{2}$ or $N_c$\\
        $SU(2)$ & $\mathbf{4}$ & 10 & 4 & Yes & 5
    \end{tabular}
    \caption{Applications of the obstruction to gauge theories with various gauge groups $G$ and matter representations $R$. The final column lists the minimum number of vacua implied by our obstruction if the theory is gapped. In the case with symmetric or antisymmetric tensor representations, the number of enforced vacua is $\frac{N_c}{2}$ or $N_c$ depending on if $N_c$ is 0 modulo 4 or not.}
    \label{tab:obst}
\end{table}

In particular, adjoint QCD has the obstruction. Indeed, when $N_f=1$ this is $\mathcal{N}=1$ super-Yang-Mills theory, and it is well known \cite{Witten:1982df} that at long distances the theory is confining and gapped with $N_{c}$ vacua arising from discrete chiral symmetry breaking due to gaugino condensation. 

The case $N_f=2$ has also been recently studied \cite{Anber:2018tcj, Cordova:2018acb, Bi:2018xvr} (see also \cite{Unsal:2007vu, Unsal:2007jx} for foundational work). The most plausible IR phase consists of $N_{c}$ vacua each supporting a $\mathbb{CP}^1$ sigma model.  Although this phase is gapless, the chiral $\mathbb{Z}_{4N_{c}}$ symmetry acts trivially on the $\mathbb{CP}^1$. Therefore, without additional massless degrees of freedom, the anomaly of the discrete chiral symmetry must be matched by spontaneous symmetry breaking and the discrete vacuum degeneracy is inevitable.

In some examples, one can realize spontaneous symmetry breaking of the discrete chiral symmetry semi-classically.   We add to the theory to a gauge-neutral complex scalar $\phi$ and include a Yukawa coupling $\phi \psi \psi$, where we assume that a gauge-neutral bilinear of $\psi$ exists \footnote{This means that $R$ is pseudoreal and $N_f$ is even, or $R$ is real.}. To preserve the $\mathbb{Z}_{N_f I(R)}$ symmetry, we let $\phi$ transform as $\phi \to e^{\frac{-4\pi \mathrm{i}}{N_fI(R)}}\phi$.  

We now spontaneously break the discrete symmetry via a potential condensing the scalar $\phi$.  Classically the phase  $\theta$ of $\phi$ is a periodic massless scalar.  However, gauge instantons generate a potential for $\theta$ through the Peccei-Quinn mechanism \cite{Peccei:1977hh}.  This potential is $\frac{2\pi}{N_fI(R)}$ periodic, leaving $N_fI(R)$ vacua breaking $\mathbb{Z}_{N_fI(R)}$ down to $\mathbb{Z}_2$. 

\paragraph{Weyl Semimetal} The anomaly obstruction derived above also has applications to systems of interest in condensed matter physics.  Here we highlight one example.  

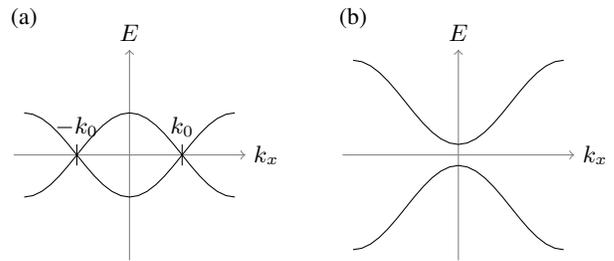
\begin{figure}[t]
    \centering
    \begin{tikzpicture}[domain=-1:1,scale = 1.4]
        \draw[->,color=gray] (-1.1,0) -- (1.1,0) node[right,black] {$k_x$};
        \draw[->,color=gray] (0,-1) -- (0,1) node[above,black] {$E$};
        \draw (-.5,-.1) -- (-.5,.1);
        \node[anchor = south] at (-.5,.1) {$-k_0$};
        \draw (.5,-.1) -- (.5,.1);
        \node[anchor = south] at (.5,.1) {$k_0$};
        \draw plot(\x,{.4*cos(3.1415*\x r)});
        \draw plot(\x,{-.4*cos(3.1415*\x r)});
        \node at (-1,1.3) {(a)};
    \end{tikzpicture}
    \hspace{.05\linewidth}
    \begin{tikzpicture}[domain=-1:1,scale = 1.4]
        \draw[->,color=gray] (-1.1,0) -- (1.1,0) node[right,black] {$k_x$};
        \draw[->,color=gray] (0,-1) -- (0,1) node[above,black] {$E$};
        \draw plot(\x,{.4*cos(3.1415*\x r)-.5});
        \draw plot(\x,{-.4*cos(3.1415*\x r)+.5});
        \node at (-1,1.3) {(b)};
    \end{tikzpicture}
    \caption{The schematic picture of the band structure \eqref{eq:Bloch}. (a) When $-1\le\xi\le$, the band has two crossings at $k_x = \pm k_0$ with energy $E=0$ which leads to the effective relativistic Weyl fermion description. (b) When $|\xi|>1$ the system is trivially gapped when half-filled.}
    \label{fig:band}
\end{figure}

In materials called Weyl semimetals \cite{YangQHall,BurkovWeyl,ChoBi2Se3,YoungDirac}, the low energy physics is effectively described by two Weyl fermions. A simple model of a Weyl semimetal is realized by the Bloch Hamiltonian acting on two-component functions of $\vec{k}$ \cite{YangQHall}
\begin{eqnarray}
    \mathcal{H}_\mathbf{k} &=&   2t_y\sin k_y\sigma_y + 2 t_z \sin k_z \sigma_z   \label{eq:Bloch}\\
    &+&\left( 2 t_x (\cos k_x - \xi) +m (2-\cos  k_y-\cos k_z)\right)\sigma_x~, \nonumber
\end{eqnarray}
where $t_y,t_z,\xi$ adn $m$ are parameters and $\sigma_i$ are the Pauli matrices.
When $-1 \le \xi \le 1$, there are band crossings at $k_x = \pm k_0 = \pm \cos^{-1}(\xi)$ with energy $E=0$ (see Fig.~\ref{fig:band}).
The dispersion around the crossings is approximately linear and thus, when half-filled, the low energy physics can be described by relativistic left-handed and right-handed Weyl fermions $\psi_1$ and $\overline{\psi}_2$ corresponding to the crossing points $k_x = \pm k_0$.  The electromagnetic $U(1)_\mathrm{EM}$ is a global symmetry of this theory and acts on the Bloch wave function as a uniform phase rotation. This in turn acts on the effective relativistic fermion fields as $\psi_1 \to e^{\mathrm{i}\alpha}\psi_1$ and $\overline{\psi}_2\to e^{\mathrm{i}\alpha}\overline{\psi}_2$ with phase $\alpha.$

On the other hand, the discrete translation $T_x$ along the $x$ direction acts on the effective fermions as the axial rotation $\psi_{1}\to e^{\mathrm{i} k_0} \psi_1$ and $\overline{\psi}_{2}\to e^{-\mathrm{i} k_0} \overline{\psi}_2$, because at a general point on the Brillouin zone it acts on Bloch wave function as multiplication of $e^{\mathrm{i} k_x}$.  When the parameters in \eqref{eq:Bloch} are tuned to special values, the location $k_{0}$ of the crossing occurs at $\frac{2\pi}{N}$ for some integer $N$.  In this case, the $T_{x}$ translations give rise to a $\mathbb{Z}_{N}$ axial symmetry.

When the truncation occurs, the effective theory has a $\mathbb{Z}_N$-$U(1)_{\text{EM}}$ mixed anomaly discussed in \cite{Cho:2017fgz}, as well as the $\mathbb{Z}_N$-gravity mixed anomaly extensively discussed above.  As remarked in 
\cite{Cho:2017fgz}, these anomalies exist only in the IR effective theory and not in the UV lattice system, and therefore the gapless modes are not protected under arbitrary continuous deformations of the Hamiltonian.  Indeed, the system can easily be gapped out by setting $|\xi|>1$, see Fig.~\ref{fig:band}(b).

The anomalies, however, can constrain the dynamics when the Lorentz symmetry and truncation of the translation symmetry to $\mathbb{Z}_N$ in the IR are preserved.  In particular, our main results says that, under the assumption above, the $\mathbb{Z}_N$-gravity anomaly protects the gapless modes even if general interactions are included \footnote{Meanwhile, the $\mathbb{Z}_N$-$U(1)_{\text{EM}}$ anomaly can be matched by a symmetry preserving topological $\mathbb{Z}_N$ gauge theory \cite{Wang:2017loc}.}.

It would be interesting to find applications of our obstruction to lattice theories with crystallographic symmetry.  For instance, in a model with point group containing $\mathbb{Z}_N$, if the long-distance physics is relativistic and realizes the anomaly the only way to gap the system while preserving the point group would be to violate the effective Lorentz symmetry.

\paragraph{Acknowledgements} We thank K. Sackel for discussions.
\bibliography{refs}

\end{document}